\documentclass[aps,prb,superscriptaddress,reprint,10pt,floatfix,showpacs]{revtex4-1}

\usepackage[utf8]{inputenc}
\usepackage{amssymb}
\usepackage{amsmath}
\usepackage{multirow}

\usepackage[pdftex]{graphicx}

\usepackage{dsfont}
\usepackage{bm}
\usepackage[normalem]{ulem}
\usepackage{hyperref}
\hypersetup{
    colorlinks,%
    citecolor=blue,%
    filecolor=blue,%
    linkcolor=blue,%
    urlcolor=blue
}

\providecommand{\ket}[1]{\mbox{\ensuremath{\vert #1\rangle}}} 

\usepackage[table]{xcolor} 
\usepackage{color}

\begin{document}
\title{Topological nonsymmorphic ribbons out of symmorphic bulk}

\author{Augusto L. Araújo}
\affiliation{Instituto de Física, Universidade Federal de Uberlândia, Uberlândia, Minas Gerais 38400-902, Brasil}

\author{Ernesto O. Wrasse}
\affiliation{Universidade Tecnológica Federal do Paraná, Toledo, Paraná 85902-040, Brasil}

\author{Gerson J.~Ferreira}
\author{Tome M. Schmidt}
\affiliation{Instituto de Física, Universidade Federal de Uberlândia, Uberlândia, Minas Gerais 38400-902, Brasil}
\date{\today}

\begin{abstract}
States of matter with nontrivial topology have been classified by their bulk symmetry properties. However, by cutting the topological insulator into ribbons, the symmetry of the system is reduced. By constructing effective Hamiltonians containing the proper symmetry of the ribbon, we find that the nature of topological states is dependent on the reduced symmetry of the ribbon and the appropriate boundary conditions. We apply our model to the recently discovered two-dimensional topological crystalline insulators composed by IV-VI monolayers, where we verify that the edge terminations play a major role on the Dirac crossings. Particularly, we find that some bulk cuts lead to nonsymmorphic ribbons, even though the bulk material is symmorphic. The nonsymmorphism yields a new topological protection, where the Dirac cone is preserved for arbitrary ribbon width. The effective Hamiltonians are in good agreement with {\it ab initio} calculations.
\end{abstract}
\pacs{73.22-f,73.20.-r}
\maketitle

\section{Introduction}

Topological insulator (TI) materials are characterized by a bulk gap with band inversions and metallic states on the borders. 
These edge (surface) states are topologically protected by symmetry.
A class of TIs protected by time-reversal symmetry has been predicted and realized experimentally
\cite{Kane2005,Bernevig2006,Konig_Science2007,HsiehNat2008,RevMPhys82-3045-2010},
where a $Z_2$ topological invariant has been used to characterize them.
The crystal lattice symmetry can also lead to topological protection on the topological crystalline insulators (TCIs)
\cite{Fu2011,HsiehNC-2012,Tanaka2012,Dziawa2012,Okada2013,ReviewAndo-Fu-2015,Xu2012,ozawa2014topological}, where the topological nontrivial states are characterized by a crystal symmetry Chern number.
Recently, it has been shown that yet a new class of topological nonsymmorphic crystalline insulators \cite{parameswaran2013topological,liu2014topological,shiozaki2015z,fang2015new,Varjas2015NonSym,Kane2015Nonsymm,chen2015topological} presenting unique properties with respect to the topological order exists.
The overall classification of topological insulators has been discussed based on space group symmetry of the bulk systems
\cite{Kane2015Nonsymm,Slager-NatPhys,fang2012bulk,jadaun2013topological}.
However, by reducing the symmetry, forming surfaces or edges, a question arises -- are the topological protected edge states completely described only by the parent bulk symmetry?

\begin{figure}[ht!]
  \centering
  \includegraphics[width=\columnwidth,keepaspectratio=true]{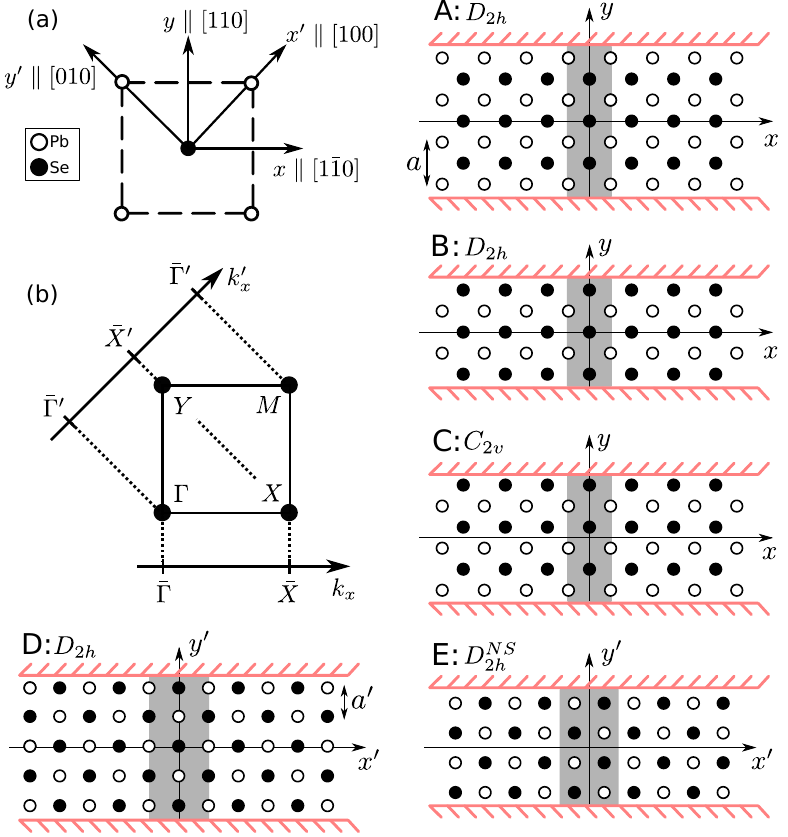}
  \caption{(a) Unit cell of the PbSe monolayer. (b) The first Brillouin zone (BZ) with the $\Gamma$, X, Y, and M TRIMs. Confinement (in red) along $y$ ($y'$) projects the TRIMs onto $k_x$ ($k_x'$) to form the one-dimensional BZ given by $\bar{X}$ and $\bar{\Gamma}$ ($\bar{X}'$ and $\bar{\Gamma}'$).
  (A-E) Illustration of the five possible crystallographic terminations of PbSe ribbons.
  The unit cell of each ribbon is highlighted in gray. The point group symmetries are indicated with usual notation, except for panel E, where ``NS'' stands for nonsymmorphic.
  The ribbons considered here have about 45 atoms ($\sim 10$~nm) along $y$ ($y'$).
  }
  \label{fig:StripeLattices}
\end{figure}

In this paper we build effective Hamiltonians using group theory \cite{voon2009kp, dresselhaus2007group, car1976optimal, koster2012space, FazzioTeoGrupos} for a TCI monolayer and ribbons to investigate the effects of the edge terminations on its topological properties. A PbSe monolayer is chosen as a representative two-dimensional (2D) TCI for our discussion \cite{Wrasse2014PbSeTCI,Fu-NL-2015,PRB91-201401-15,kobayashi2015electronic}. The band structures from the effective Hamiltonians are compared with {\it ab initio} results obtained from density functional theory (DFT) calculations using the {\sc VASP} code \cite{kresse1996efficient}. We consider five possible crystallographic ribbon cuts, starting from the simpler case A (Fig.~\ref{fig:StripeLattices}) building up in complexity towards our main result in ribbon E. We find that the energy dispersion of the topological edge states is strongly dependent on both the reduced symmetry and boundary conditions, resembling graphene's zigzag and armchair edges \cite{divincenzo1984graphene,BreyFertig2006Ribbons}. Interestingly, we show that while our bulk monolayer is a symmorphic lattice, one particular cut leads the nonsymmorphic ribbon E, whose symmetry group is not a subgroup of its bulk counterpart. It is known that nonsymmorphism yields extra degenerescences with respect to its underlying point group \cite{voon2009kp, dresselhaus2007group, car1976optimal, koster2012space, FazzioTeoGrupos}, which in our case results in an extra protection that preserves the Dirac cone for ribbons of arbitrary width. In addition to the fundamental physics presented here, the nonsymmorphic systems could be potential materials for nanoscale 2D devices, preserving the topological state properties even for nanosized ribbons.


Experimentally, atomic layer growth control of IV-VI materials has been achieved via electrochemical atomic layer epitaxy/deposition \cite{vaidyanathan2004quantum, vaidyanathan2006preliminary}, and PbSe nanorods and nanotubes were recently grown \cite{li2012facile}. However, a refined edge control remains as challenging as for any other 2D material. Recently developed chemical bottom-up approaches were successful in graphene \cite{narita2015bottom,jacobberger2015direct}. 
Effects of edge saturation and substrates are yet to be experimentally explored.
Recently, first-principles calculations showed that the topological properties and the energetic stability of IV-VI monolayers can be manipulated using appropriate substrate \cite{kobayashi2015electronic}.


\section{Model for the infinite monolayer}

\begin{figure}[ht!]
  \centering
  \includegraphics[width=\columnwidth,keepaspectratio=true]{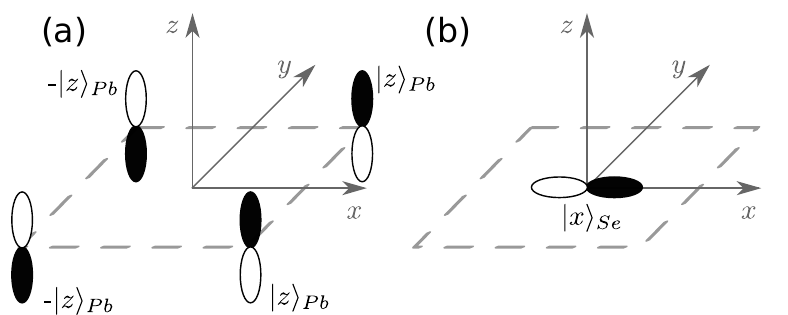}
  \caption{Representation on the unit cell of the spinless eigenstates of (a) conduction and (b) valence bands at $X$. The orientation of the $p_z$ orbitals of Pb in (a) are chosen to satisfy the Bloch periodicity at $X$, and the resulting state $\ket{xz;s}$ is odd along both $x$ and $z$ directions. The state $\ket{x;s}$ in (b) is composed by a single $p_x$ orbital of Se on the unit cell.}
  \label{fig:Lattice-BZ}
\end{figure}

The PbSe monolayer has a square Bravais lattice [Fig.~\ref{fig:StripeLattices}(a)] with a space group symmetry $D_{4h}$. The time-reversal invariant momenta (TRIMs) are $\Gamma$, X, Y, and M, as shown in Fig.~\ref{fig:StripeLattices}(b). From first principles \cite{Wrasse2014PbSeTCI} it is known that the band inversions occur at $X$ and $Y$, where the symmetry is reduced to the $D_{2h}$ space group. At X, without spin-orbit (SO) couplings the top of the valence band is composed mostly by $p_x$ orbitals of Se, while the bottom of the conduction band is dominated by $p_z$ orbitals of Pb. At Y, all properties are given by a $C_4$ rotation of X, which allow us to focus our discussion on the X point. To satisfy the Bloch theorem at X the orbitals must be arranged periodically along $y$ and anti-periodically along the $x$ direction, as shown in Fig.~\ref{fig:Lattice-BZ}. To emphasize the Cartesian symmetry of these basis states, we label the kets referring to the axes through which the state is odd. Hence, the states in Figs.~\ref{fig:Lattice-BZ}(a) and \ref{fig:Lattice-BZ}(b) become $\ket{xz;s}$ and $\ket{x;s}$, respectively. Here $s = \{\uparrow, \downarrow\}$ denotes the spin.

\begin{figure}[ht!]
  \centering
  \includegraphics[width=0.95\columnwidth,keepaspectratio=true]{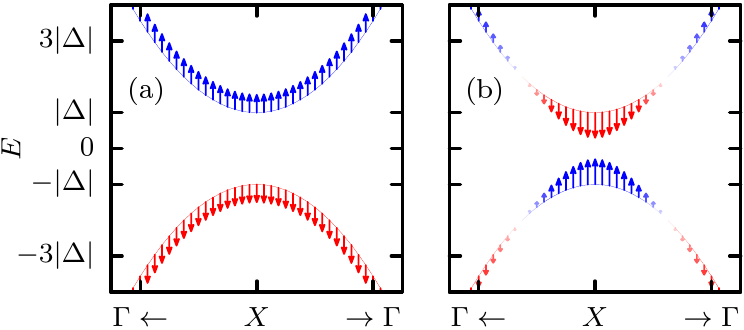}
  \caption{Band structure of the PbSe monolayer from the effective model for $k_y = 0$ and as a function of $k_x$. (a) Without SO we consider $\Delta = m_{x} = 1$, and $\delta m_x = \alpha_x = 0$. (b) With SO the gap changes sign $\Delta = -1$, the masses remain the same, and $\alpha_x = 2$. The colors represent the contributions from the $\ket{x;s}$ (red) and $\ket{xz;s}$ (blue) orbitals in accordance with the \textit{ab initio} data \cite{Wrasse2014PbSeTCI}. The arrows represent the spin projection along $z$.
  }
  \label{fig:bsX}
\end{figure}

From this set of orbitals, $\{ \ket{xz;\uparrow}, \ket{xz;\downarrow}, \ket{x;\uparrow}, \ket{x;\downarrow} \}$, we construct the effective Hamiltonian for the monolayer considering a $k$-space expansion \cite{divincenzo1984graphene}. Following this ordering, one obtains a matrix representation for the symmetry elements of the $D_{2h}$ group, plus the time-reversal operator $\mathcal{T}$ \footnote{See Appendix for the matrix representation of the symmetry elements, and character tables of space groups.}. Requiring that the Hamiltonian $H_{X}$, for $\bm{k}$ around X, commutes with these symmetry elements and $\mathcal{T}$ up to second order in $\bm{k}$, we obtain

\begin{multline}
 H_{X} = \Delta\tau_z + (\alpha_y k_y \sigma_x  - \alpha_x k_x \sigma_y)\tau_x \\
  + (m_{x}\tau_z + \delta m_x)k_x^2 +(m_{y}\tau_z + \delta m_y)k_y^2,
 \label{eq:Hmonolayer}
\end{multline}
where the Pauli matrices $\sigma_i$ ($\tau_i$) act on the spin (orbital) degrees of freedom. 
From the $k\cdot p$ theory one can associate the $\alpha_x$ and $\alpha_y$ with the $k$-dependent SO contribution, while the gap $\Delta$ has contributions both from the bare lattice potential $V(r)$ and from the $k$-independent SO term via remote bands. Here $\Delta$ plays the role of the Dirac mass and changes sign as a function of the SO intensity. The mass (parabolic) terms $m_{(x,y)}$ are anisotropic, and $\delta m_{(x,y)}$ could break the particle-hole symmetry. This Hamiltonian describes the bulk PbSe monolayer, as we can see in Fig.~\ref{fig:bsX}, where the band structure around $k=X$ without ($\Delta = 1$) and with SO couplings ($\Delta = -1$) are in qualitative agreement with {\it ab initio} data \cite{Wrasse2014PbSeTCI}. 

From the effective bulk Hamiltonian, we can calculate the topological invariants of the system, i.e., the Chern numbers. The bulk monolayer and all possible ribbons share only the identity and mirror $\mathcal{M}$ ($z\rightarrow -z$) symmetry elements. Therefore, all eigenstates belong to one of two classes \cite{ReviewAndo-Fu-2015} defined by the eigenvalues $\eta = \pm i$ of the mirror operator, i.e., $\mathcal{M} \ket{\psi_\eta} = \eta \ket{\psi_\eta}$. For each class one defines a Chern number $N_\eta$, which allows us to calculate the total Chern number $N_T = N_{+i} + N_{-i}$, and the mirror Chern number $N_\mathcal{M} = (N_{+i}-N_{-i})/2$. For $\Delta > 0$ we find all $N_\eta = 0$ and the system is on the trivial regime as expected. For $\Delta < 0$ the states from the occupied bands give $N_{\pm i} = \mp 2$, which yields $N_T = 0$ and $N_\mathcal{M} = -2$, thus characterizing the TCI phase \cite{ReviewAndo-Fu-2015}.

\section{TCI Ribbons}

By cutting the monolayer into ribbons, the introduced lateral confinement may break some symmetries of the system, allowing new terms into the effective Hamiltonian. There are two main crystallographic orientations for the ribbons: $x \parallel [1\bar{1}0]$ and $x' \parallel [100]$ directions. The first has three possible edge terminations, illustrated in Fig.~\ref{fig:StripeLattices}, panels A, B, and C, while the latter has two more possibilities, shown in Fig.~\ref{fig:StripeLattices}, panels D and E. Hereafter we refer to each termination by these capital letters. Next to each panel in Fig.~\ref{fig:StripeLattices} we label the corresponding space group symmetry.

\subsection{Ribbons A, B, and C}

Let us first consider ribbons A, B, and C, which constitute our simplest cases and provide the context to advance to more complex scenarios. First, since ribbons A and B belong to the same $D_{2h}$ space group, they must have the same Hamiltonian $H_A = H_B$.
The confinement along $y$ projects Y and $\Gamma$ into $\bar{\Gamma}$, and M and X into $\bar{\rm X}$ (see Fig.~\ref{fig:StripeLattices}). As the bulk gaps at $\Gamma$ and M are much larger than the gap at X and Y, we can neglect the extra bands coming from these projections. The resulting Brillouin zone of the ribbon is given by the $\bar{\rm X}$ and $\bar{\Gamma}$ TRIMs, whose effective Hamiltonians are obtained replacing $k_y \rightarrow -i \partial/\partial y$ in $H_X$ and $H_Y = C_4 H_X C_4^{-1}$, respectively. Both are in the TCI regime and one can expect topologically protected states at both $\bar{\Gamma}$ and $\bar{\rm X}$. 
Second, ribbon C belongs to the $C_{2v}$ space group, which is a subgroup of $D_{2h}$.
However, we find that the extra terms \cite{Note1} in the Hamiltonian for ribbon C play no significant role in the qualitative analysis of the topological properties discussed here. Hence, its effective Hamiltonian $H_C\approx H_A=H_B$. Ultimately, the distinction between ribbons A, B, and C arises from their different terminations, which enter our effective model via boundary conditions.

To establish the appropriate boundary conditions for the envelope functions for each termination, we extend Brey and Fertig's approach \cite{BreyFertig2006Ribbons} from graphene to our PbSe monolayer ribbon of width $2W$. In ribbon A the edges are composed by Pb atoms, say at $y = \pm W$. If the ribbon was uncut, the next line of atoms on top would be of Se at $y = W + a/2$, where $a$ is the lattice parameter (Fig.~\ref{fig:StripeLattices}). Following those, there would be yet a line of Pb atoms at $y = W + a$. However, since those atom lines were cut off to form the ribbon, we set the envelope function of each sublattice to zero at these positions. A similar definition follows for the bottom edge of ribbon A, and a generalization to ribbons B and C is immediate. Note that the boundary condition for the top edge of ribbon C is equivalent to those of ribbon B, while the bottom of C is equivalent to the boundaries of A. 

\begin{figure*}[ht!]
  \centering
  \includegraphics[width=\textwidth,keepaspectratio=true]{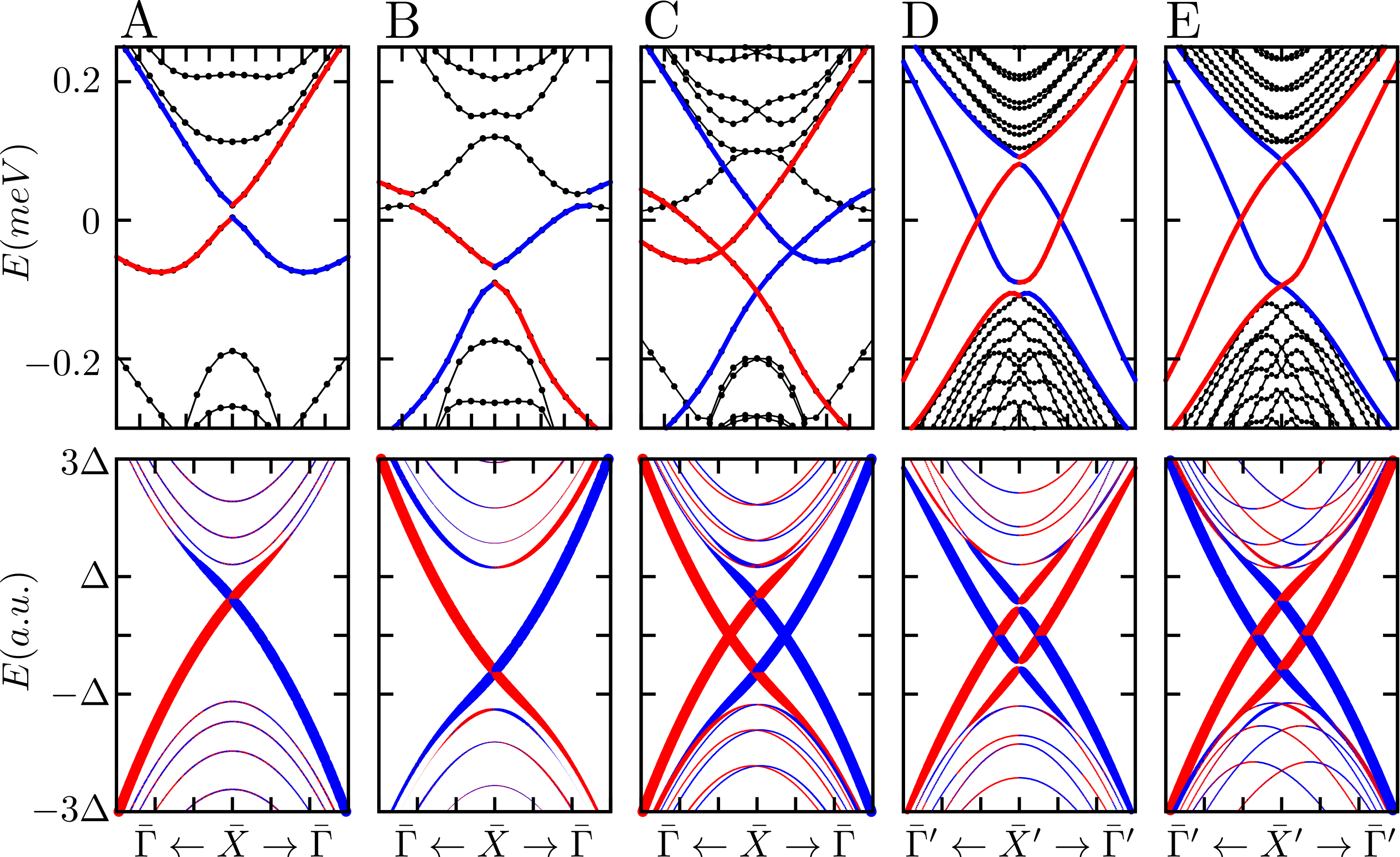}
  \caption{Comparison of the \textit{ab initio} (top) and effective model (bottom) band structures of PbSe topological insulator ribbons A, B, C, D and E. The colors represent the spinfull mirror parity $+i$ (red) and $-i$ (blue). The thickness of the lines for the effective model represents the localization of the state into an edge. In A, D, and E we show the states and mirror parities projected on the top edge of the ribbons. In B these projections are taken for the bottom edge. In C the upper (lower) Dirac crossing belongs to the top (bottom) edge states. In A, B, D, and E the states shown here are degenerate with ones from the opposite edge, with opposite mirror parity. The {\it ab initio} data are colored by hand as a guide to the eyes.}
  \label{fig:BSABCDE}
\end{figure*}

The resulting band structures of ribbons A, B, and C are shown in the bottom panels of Fig.~\ref{fig:BSABCDE} to be compared with the {\it ab initio} data on top. We can see some differences between the effective model and the {\it ab initio} results. First, we observe a band gap at top panels A and B, which is absent from our model (bottom panels). The band gap opening occurs because the ribbon in the DFT calculation is narrow, $2W \approx 10$~nm, and quantum tunnel coupling between topological states from opposite edges takes place. This hybridization gap vanishes asymptotically with the ribbon width. We also observe additional states crossing the Fermi level in the \textit{ab initio} results (upper panels of Fig.~\ref{fig:BSABCDE}, panels A-C), due to dangling bonds at the edges. 

\subsection{Ribbons D and E}

The other possible terminations occur in ribbons aligned along the $(x', y')$ coordinates, illustrated in Fig.~\ref{fig:StripeLattices}, panels D and E, having both Pb and Se atoms at the edges.
In these cases we set the envelope functions to zero on the next (absent) line of atoms of both lattices, i.e., at $|y'|=W+a'$, with $a' = a/\sqrt{2}$. Since the boundary conditions for ribbons D and E are the same, their distinction occurs only via the different space groups. Next we show that this translates into distinct XY valley couplings. Interestingly, for ribbon E, the nonsymmorphism of the crystal lattice introduces extra topological protections.

Ribbon D belongs to the $D_{2h}$ space group, the same of ribbons A and B. 
The confinement along $y'$ projects M and $\Gamma$ into $\bar{\Gamma}$ and both X and Y into $\bar{\rm X'}$ [see Fig.~\ref{fig:StripeLattices}(a)]. Therefore one can expect topological states only at $\bar{\rm X'}$. Since the monolayer bands around Y are simply the ones at X rotated by $C_4$, to obtain the effective Hamiltonian around $\bar{\rm X'}$ we have to expand our basis to incorporate the bands coming from Y. Therefore we establish a new matrix representation for the symmetry elements of $D_{2h}$ with a basis set as $\{ \ket{xz;\uparrow}, \ket{xz;\downarrow}, \ket{x;\uparrow}, \ket{x;\downarrow}, \ket{yz;\uparrow}, \ket{yz;\downarrow}, \ket{y;\uparrow}, \ket{y;\downarrow} \}$. Following the same procedure discussed previously for $H_X$, we obtain

\begin{equation}
H_{X'} = 
\begin{pmatrix}
  \tilde{H}_X & V_{XY}\\
  V_{XY}^\dagger & \tilde{H}_Y
\end{pmatrix},
\label{eq:HDE}
\end{equation}
where $\tilde{H}_X$ and $\tilde{H}_Y$ are equivalent to $H_X$ and $H_Y$ rotated towards the $\bm{r}'$ coordinates, i.e., $\hat{x} \rightarrow (\hat{x}'-\hat{y}')/\sqrt{2}$, $\hat{y} \rightarrow (\hat{x}'+\hat{y}')/\sqrt{2}$, and equivalent rotations for $k_x$, $k_y$, $\sigma_x$ and $\sigma_y$. The relevant term in the XY valley hybridization $V_{XY}$ for ribbon D is

\begin{equation}
V_{XY} \approx i v_D k_y^2\tau_z\sigma_z.
\label{eq:vxyD}
\end{equation}

If $V_{XY}$ were zero, the boundary conditions of ribbon D would give us four degenerate Dirac cones at $E=0$, leading to eight-fold degeneracy (labeled by $\mathcal{M} = \pm i$, X/Y valleys, and top/bottom edges). However, the main contribution to $V_{XY}$ shown in Eq.~\eqref{eq:vxyD}, which is proportional to the mirror symmetry operator $\mathcal{M} = i\tau_z\sigma_z$, couples states from opposite valleys with the same mirror eigenvalue ($\pm i$). This splits the Dirac points into the top and bottom (gapped) cones of Fig.~\ref{fig:BSABCDE}, panel D. The hybridization gap opens due to the coupling between top and bottom edge states, which is allowed by Eq.~\eqref{eq:vxyD}, and consistent with the group character tables shown in the Appendices, which guarantees only two-fold degeneracy for ribbon D. Other terms of $V_{XY}$ are shown in the Appendices. Their contribution is only quantitative to the fine tuning of the band structure.

Our main result is the unexpected nonsymmorphic space group $D_{2h}^{NS}$ of ribbon E, which yields extra topological protections \cite{dresselhaus2007group,Kane2015Nonsymm}.
Here the point group symmetry elements are the same as in $D_{2h}$, however, some of them must be complemented by a nonprimitive translation $\bm{\chi} = a'\hat{x}'$ of half a unit cell along $x'$, i.e., glide planes and screw axis elements \cite{voon2009kp, dresselhaus2007group, car1976optimal, koster2012space, FazzioTeoGrupos}. To obtain the matrix representation for these symmetry elements we use the same basis set from ribbon D above, but with the coordinates shifted (see Fig.~\ref{fig:StripeLattices}).
Requiring that the Hamiltonian commutes with these elements and $\mathcal{T}$, we obtain again $H_{X'}$, but with different XY valley couplings $V_{XY}$, whose relevant terms for ribbon E are

\begin{equation}
V_{XY} \approx v_E k_y^2(i\tau_x \sigma_y - \tau_y \sigma_x).
\end{equation}

Similarly to ribbon D, this XY valley coupling splits the otherwise eight-fold degeneracy into the top and bottom Dirac cones of Fig.~\ref{fig:BSABCDE}, panel E. Here they remain gapless. Other terms of $V_{XY}$ compatible with the symmetries of ribbon E contribute only with the fine tuning of the band structure. These are shown in the Appendices.

We emphasize that the edge state dispersions for ribbons D and E, shown in Fig.~\ref{fig:BSABCDE}, panels D-E, differ only by the hybridization gaps at X, which is the main consequence of the nonsymmorphic lattice of ribbon E.
Here, all edge state branches are doubly degenerate with states located at opposite edges having opposite mirror parities. The color code in the figure refers to the states located on the top edge. Therefore, at each crossing around X$'$ there are four states. The agreement between the effective Hamiltonian and the {\it ab initio} results is patent. 
In ribbon D the gap between edge branches is a consequence of top/bottom edge hybridization for narrow ribbons. The gap is closed in the nonsymmorphic ribbon E, which is consistent with its double group character table\cite{Note1}. We have calculated this character table using standard group theory method \cite{dresselhaus2007group} to find that the double group is composed by two-dimensional irreducible representations (IRREPs), which yields two-fold degeneracies. However, while for ribbons A, B, C, and D the time-reversal symmetry does not lead to extra degeneracies, for ribbon E the pair of 2D IRREPs form Kramers partners. This leads to the four-fold degeneracy of the edge states of ribbon E at X$'$, which \textit{must survive even for narrow ribbons or arbitrary width, despite the overlap between top and bottom edge states}. Usual edge state branches, as in ribbon D, can only close the gap asymptotically for wide ribbons, constituting an accidental degeneracy, which is not protected by symmetry.

\section{Final remarks and conclusions}

The double crossing band structure of ribbon C resembles those of D and E, with closed gaps. However, here edge state branches are nondegenerate. The bottom crossing at X corresponds to a pair of opposite mirror parity states coming from the top edge, which are equivalent to half of the edge states of ribbon B. Similarly the top crossing at X involves states from the bottom edge of ribbon C, which are equivalent to half of the states of ribbon A. 
Consequently, the crossings of pairs of edge states from opposite sides are already split in energy, thus avoiding a direct hybridization, which keeps the gap closed, in contrast to its counterparts in ribbons A and B. 

Interestingly, the gap oscillation with even/odd number of layers recently reported in Ref.~\onlinecite{ozawa2014topological} originates in the alternation between symmorphic and nonsymmorphic lattices in three dimensions, which can be understood as a 3D counterpart of our results.

In conclusion, we showed that, although the Chern number and topological classification of insulators remain a bulk property, the reduced symmetry of the ribbons and the characteristics of its edge terminations play a fundamental role in the topological state properties. Particularly, we focused on the TCIs given by IV-VI monolayers, whose pair of distinct atoms lead to distinct sublattices, similarly to graphene. While different cuts of the bulk in graphene give us the armchair and zigzag ribbons, here we identify five main types of ribbons due to the more complex structure of the lattice. Interestingly, we find that the extra topological protection introduced by nonsymmorphic symmetry yields protected crossings for ribbons of arbitrary width. This is in contrast with the usual topological protections, where the gaps are only asymptotically closed for large enough samples. The extra nonsymmorphic protection of the ribbon could not be predicted by the bulk topological classification, since the bulk is symmorphic. This feature may allow topological properties to be explored in nanoscale nonsymmorphic TIs.

\begin{acknowledgments}
The authors acknowledge the financial support from the Brazilian Agencies CNPq, CAPES, and FAPEMIG, and the computational facilities from CENAPAD.
\end{acknowledgments}

\appendix 

\section{Matrix Representations and Effective Hamiltonians}

In order to obtain the effective Hamiltonian for each PbSe ribbon via the method of invariants\cite{voon2009kp,dresselhaus2007group}, we define a matrix representation for each symmetry element of the space group of the ribbon. Then, imposing that the effective Hamiltonian must commute with these symmetry elements and the time-reversal operator, we find the allowed finite elements of the effective Hamiltonian. Below we present these matrix representations for each ribbon. Except for ribbon E, in all other ribbons the space group matches the point group of the unit cell plus primitive translations. Therefore we label the space groups with usual Schönflies notation of its corresponding point group. Ribbon E is a particularly interesting case, as it belongs to a nonsymmorphic space group\cite{car1976optimal, koster2012space, dresselhaus2007group}, and its symmetry operations are a combination of point group symmetries and translations by half a unit cell, i.e. glide planes and screw axes.

\subsection{Monolayer Sheet and Ribbons A and B}
\label{app:monolayer}

The full PbSe monolayer belongs to the $D_{4h}$ space group. However, the space group for $\bm{k} = X$ point is $D_{2h}$. For ribbons A and B, both the lattice and the $\bm{k} = X$ point belong to the $D_{2h}$ as well. Therefore, the effective Hamiltonians around $\bm{k}=X$ for the full monolayer ($H_X$) and for ribbons A ($H_A$) and B ($H_B$) are essentially the same, i.e. $H_A = H_B = H_X$. The distinction between these cases arises only via the boundary conditions. For the full monolayer sheet there is no boundaries and both $k_x$ and $k_y$ remain good quantum numbers. For ribbons A and B, the confinement along $y$ requires $k_y \rightarrow -i \partial/\partial y$, and the different boundary conditions for A and B are discussed in the main text of this paper.

Let us consider only the effective Hamiltonian around $\bm{k} = X$, which must commute with the symmetry operations of $D_{2h}$: identity $E$; two-fold rotations around the $x$, $y$ and $z$ axes, $C_2(x)$, $C_2(y)$, and $C_2(z)$; inversion $I$, and mirror planes $M(xy)$, $M(xz)$, $M(yz)$. From the \textit{ab-initio} data, we know that the relevant orbitals around the Fermi level are $\{ \ket{xz;\uparrow}, \ket{xz;\downarrow}, \ket{x;\uparrow}, \ket{x;\downarrow} \}$. These orbitals are shown in Fig.~\ref{fig:Lattice-BZ}, from which we obtain by inspection the matrix representation of the symmetry operations as

\begin{align}
 \nonumber  
 E =& \tau_0 \otimes \sigma_0,\\
 \nonumber
 C_2(z) =& -\tau_0 \otimes (-i \sigma_z),\\
 \nonumber
 C_2(x) =& -\tau_z \otimes (-i \sigma_x),\\
 \nonumber
 C_2(y) =& \tau_z \otimes (-i \sigma_y),\\
 I =& \tau_z \otimes \sigma_0,\\
 \nonumber \label{eq:mirrorAB}
 M(xy) =& -\tau_z \otimes (-i \sigma_z),\\
 \nonumber
 M(xz) =& \tau_0 \otimes (i \sigma_y),\\
 \nonumber
 M(yz) =& -\tau_0 \otimes (-i \sigma_x),
\end{align}
where $\sigma_0$ and $\tau_0$ are $2\times2$ identity matrices, $\sigma_{x,y,z}$ are the Pauli matrices acting on spin space, $\tau_{x,y,z}$ are the Pauli matrices acting on the orbitals, and $\otimes$ is the Kronecker product. Additionally, the time-reversal operator is $\mathcal{T} = \tau_0\otimes(i\sigma_y)\mathcal{K}$, where $\mathcal{K}$ is the complex conjugation.

To find the effective Hamiltonian, we start with the most general expression up to second order in $k_x$ and $k_y$,

\begin{equation}
 H_{X} = \sum_{i=0}^2 \sum_{j=0}^2 Q_{i,j} k_x^i k_y^j,
 \label{eq:HQ}
\end{equation}
where each $Q_{i,j}$ is a general Hermitian $4\times 4$ matrix, and $\bm{k}$ is deviation from the $X$ point of the BZ. Requiring that $H_{X}$ commutes with all the symmetry elements of $D_{2h}$ and $\mathcal{T}$, we obtain the effective Hamiltonian for the full monolayer and ribbons A and B, Eq.~\eqref{eq:Hmonolayer}. 

Note that in order to evaluate the commutation of $H_X$ with the symmetry elements, one must consider also the action of the symmetry operation in $k_x$ and $k_y$. For instance, taking the term $i=0$, $j=1$ of the sum above, and the symmetry operation $M(xz)$, we have

\begin{align}
\nonumber
 [Q_{0,1}k_y, M(xz)] =& Q_{0,1}k_y M(xz)- M(xz)Q_{0,1}k_y\\
 \nonumber
 =& -Q_{0,1}M(xz)k_y - M(xz)Q_{0,1}k_y \\
 =& -\{Q_{0,1}, M(xz)\}k_y,
\end{align}
where $[p,q]$ and $\{p,q\}$ are the commutator and anti-commutator of $p$ and $q$. On the second line above we have used $\{k_y, M(xz)\} = 0$, since $M(xz)$ makes $y \rightarrow -y$. Similar considerations must be taken carefully for all other terms of the sum and symmetry operations.

\subsection{Ribbon C}
\label{app:ribbonC}

To obtain the effective Hamiltonian for ribbon C we follow the same procedure presented above for ribbons A and B. However, both the lattice and the $\bm{k} = X$ point of ribbon C belong to the $C_{2v}$ space group, which is composed the by the symmetry elements $E$, $C_2(y)$, $M(xy)$, and $M(yz)$, \textit{i.e.}, a subgroup of $D_{2h}$. Consequently, we can use the same matrix representations as above. Due to the reduced symmetry of $C_{2v}$, the effective Hamiltonian for C is given by $H_X$ above, plus extra allowed terms,

\begin{equation}
\begin{split}
 H_C =& H_X + (\gamma_0 + \gamma_1 \tau_z) k_x \sigma_z + \Delta_1 \tau_y \sigma_x\\
      &+ m_{1x} k_x^2 \tau_y\sigma_x + m_{1y} k_y^2 \tau_y\sigma_x\\
      &+ \beta k_x k_y \tau_y\sigma_y.
\end{split}
\end{equation}

These extra terms contribute to the fine tunning of band structure. However, they are not relevant for the qualitatively analysis of the topological edge states presented in the main text, therefore we choose to set $H_C \approx H_X = H_A = H_B$. In Fig.~\ref{fig:BSABCDE}(A-C) we show the band structure of ribbons A, B and C, all taken from the same $H_X$. The results differ only by the boundary conditions applied in each case, showing already a good qualitatively agreement with the \textit{ab-initio} data.

\subsection{Ribbon D}

In Fig.~\ref{fig:StripeLattices}(D) we have again the point group $D_{2h}$. However, here the lattice is aligned along the $\bm{r}' = (x',y')$ coordinates. In this case, the confinement along $y'$ projects both $X$ and $Y$ TRIMs into $X'$, doubling the number of relevant bands near the Fermi level. Since $Y$ is equivalent to $X$ rotated by $C_4(z)$, the new set of relevant states are $\{ \ket{xz;\uparrow}, \ket{xz;\downarrow}, \ket{x;\uparrow}, \ket{x;\downarrow}, \ket{yz;\uparrow}, \ket{yz;\downarrow}, \ket{y;\uparrow}, \ket{y;\downarrow} \}$. The last four states there are simply those of Fig.~\ref{fig:Lattice-BZ} rotated by $C_4(z)$. From these states, the symmetry operations take the form of  $8\times 8$ matrices,

\begin{align}
 \nonumber
 E &= \Lambda_{00} \otimes \sigma_0,\\
 \nonumber
 C_2(z') &= -\Lambda_{00} \otimes (-i \sigma_z),\\
 \nonumber
 C_2(x') &= \Lambda_{zx} \otimes (-i \sigma_x),\\
 \nonumber
 C_2(y') &= -\Lambda_{zx} \otimes (-i \sigma_y),\\
 I &= \Lambda_{z0} \otimes \sigma_0,\\
 \nonumber
 M(x'y') &= -\Lambda_{z0} \otimes (-i \sigma_z),\\
 \nonumber
 M(x'z') &= -\Lambda_{0x} \otimes (i \sigma_y),\\
 \nonumber
 M(y'z') &= \Lambda_{0x} \otimes (-i \sigma_x).
\end{align}
The $4\times 4$  $\Lambda$-matrices act on the orbitals from both $X$ and $Y$ valleys, while the \mbox{$\sigma$-matrices} act on spin. The $\Lambda$-matrices are

\begin{align}
\nonumber
 \Lambda_{00} =&
 \begin{pmatrix}
   \tau_0 & 0\\
   0 & \tau_0
 \end{pmatrix},
&
 \Lambda_{zx} =&
 \begin{pmatrix}
   0 & \tau_z\\
   \tau_z & 0
 \end{pmatrix},
\\
 \Lambda_{0x} =&
 \begin{pmatrix}
   0 & \tau_0\\
   \tau_0 & 0
 \end{pmatrix},
&
 \Lambda_{z0} =&
 \begin{pmatrix}
   \tau_z & 0\\
   0 & \tau_z
 \end{pmatrix},
\end{align}
More generally, $\Lambda_{\mu\nu} = \lambda_\nu \otimes \tau_\mu$,
where $\lambda_0$ is the $2\times 2$ identity matrix, and $\lambda_{(x,y,z)}$ are Pauli matrices acting on the $\{X, Y\}$ valley subspace.
Now the time-reversal operator is \mbox{$\mathcal{T} = \Lambda_{00} \otimes (i\sigma_y)\mathcal{K}$}.

To find the effective Hamiltonian $H_{D}$ we consider again an expansion of the form of Eq.~\eqref{eq:HQ}, but now the $Q_{i,j}$ are general Hermitian $8\times 8$ matrices. Requiring that $H_{D}$ commutes with the symmetry elements above, we obtain the $H_{D}$ shown in Eq.~\eqref{eq:HDE}, with

\begin{equation}
\label{eq:fullVXYD}
\begin{split}
 V_{XY} =& (V_{r}+\delta V_{r}\tau_z) + i(V_{i}+\delta V_{i}\tau_z)\sigma_z\\
         +& (w_{rx}+\delta w_{rx}\tau_z)k_{x'}^2 + i(w_{ix}+\delta w_{ix}\tau_z)k_{x'}^2\sigma_z\\
         +& (w_{ry}+\delta w_{ry}\tau_z)k_{y'}^2 + i(w_{iy}+\delta w_{iy}\tau_z)k_{y'}^2\sigma_z\\
         +& (i \beta_{ix} \tau_y \sigma_x - \beta_{rx} \tau_x\sigma_y)k_{x'}\\
         +& (-i \beta_{iy} \tau_y \sigma_y - \beta_{ry} \tau_x\sigma_x)k_{y'}.
\end{split}
\end{equation}
Most of these terms of the $XY$ intervalley coupling contribute only to the fine tuning of the band structure. In the main text we show only the term that is relevant for the qualitatively analysis of the topological edge states.

\subsection{Ribbon E}

Ribbon E in Fig.~\ref{fig:StripeLattices}(E) belongs to a nonsymmorphic symmetry group, which we label as $D_{2h}^{NS}$. Some of the symmetry elements require a translation $\chi$ by half a unit cell along $x'$, i.e. glide planes and screw axes. Using the standard notation\cite{car1976optimal,dresselhaus2007group}, the symmetry elements are $\{E,0\}$, $\{C_2(z'),0\}$, $\{C_2(x'),\chi\}$, $\{C_2(y'),\chi\}$, $\{I,0\}$, $\{M(x'y'),0\}$, $\{M(x'z'),\chi\}$, $\{M(y'z'),\chi\}$. Using the same set of states as for Ribbon D, by inspection we find the matrix representation for these operations as

\begin{align}
 \nonumber
 \{E,0\} =& \Lambda_{0,0}\otimes \sigma_0,\\
 \nonumber
 \{C_2(z'),0\} =& \Lambda_{0,z}\otimes (-i \sigma_z),\\
 \nonumber
 \{C_2(x'),\chi\} =& -i\Lambda_{z,y}\otimes (-i \sigma_x),\\
 \nonumber
 \{C_2(y'),\chi\} =& \Lambda_{z,x}\otimes (-i \sigma_y),\\
 \label{eq:NSelements}
 \{I,0\} =& -\Lambda_{z,z}\otimes \sigma_0,\\
 \nonumber
 \{M(x'y'),0\} =& -\Lambda_{z,0}\otimes (-i \sigma_z),\\
 \nonumber
 \{M(x'z'),\chi\} =& i\Lambda_{0,y}\otimes (i \sigma_y),\\
 \nonumber
 \{M(y'z'),\chi\} =& -\Lambda_{0,x}\otimes (-i \sigma_x),
\end{align}

Once again, to find $H_E$ we require it to commute with all symmetry elements above and the time-reversal operator $\mathcal{T} = \Lambda_{00} \otimes (i\sigma_y)\mathcal{K}$. The resulting effective Hamiltonian is written again as in Eq.~\eqref{eq:HDE}, but with a different $XY$ inter-valley coupling

\begin{equation}
\label{eq:fullVXYE}
\begin{split}
  V_{XY} =& -d_0 \tau_y \sigma_x + i d_1 \tau_x\sigma_y\\
         +& (a_1 + \delta a_1\tau_z \sigma_z)k_{x'} + i(a_2 + \delta a_2 \tau_z)k_{x'}\\
         +& (-w_{0x}\tau_y\sigma_x + i w_{1x} \tau_x\sigma_y)k_{x'}^2\\
         +& (-w_{0y}\tau_y\sigma_x + i w_{1y} \tau_x\sigma_y)k_{y'}^2\\
         +& (-w_{xy0}\tau_y\sigma_y + i w_{xy1}\tau_x\sigma_x)k_{x'}k_{y'}
\end{split}
\end{equation}
Once again, in the main text we show only the term relevant for the qualitatively analysis of the topological edge states.

\section{Character Tables, Irreducible Representations, and Degeneracies}
\label{app:tables}

\subsection{Ribbons A, B, C, and D}

The character tables of the space groups of ribbons A, B, C, and D are simply those of their equivalent point groups. Those can be found in Koster's book\cite{koster2012space}, and we reproduce them here for convenience. The $D_{2h}$ group of the monolayer sheet and ribbons A, B, and D is shown in Table \ref{tab:D2h}, and the $C_{2v}$ group of ribbon C is shown in Table \ref{tab:C2v}. 

\begin{table}
 \caption{Character Table for the $D_{2h}$ double group. The first line labels the symmetry operations in short notation. For the full monolayer and ribbons A and B we have 
 $C_2 = C_2(z)$, $C'_2 = C_2(x)$, $C''_2 = C_2(y)$, $M = M(xy)$, $M' = M(xz)$, and $M''=M(yz)$. For ribbon D simply replace the coordinates $(x,y,z)$ by $(x', y', z')$. Barred and unbarred operations refer to $4\pi$ and $2\pi$ rotations on spin space. The first set, from $\Gamma_0$ to $\Gamma_7$, are single group irreducible representations, and $\Gamma_8$ and $\Gamma_9$ are the double group irreducible representations.}
 \label{tab:D2h}
\begin{center}
\begin{tabular}{l | c c c c c c c c c c}
\hline\hline
 & $E$ & $\bar{E}$ & $\begin{array}{c}C_2 \\ \bar{C}_2\end{array}$ & $\begin{array}{c}C'_2 \\ \bar{C}'_2\end{array}$ & $\begin{array}{c}C''_2 \\ \bar{C}''_2\end{array}$ & $I$ & $\bar{I}$ & $\begin{array}{c}M \\ \bar{M}\end{array}$ & $\begin{array}{c}M' \\ \bar{M}'\end{array}$ & $\begin{array}{c}M'' \\ \bar{M}''\end{array}$\\
\hline
$\Gamma_0$ & 1 & 1 &  1 &  1 &  1 &  1 &  1 &  1 &  1 &  1 \\
$\Gamma_1$ & 1 & 1 & -1 &  1 & -1 &  1 &  1 & -1 &  1 & -1 \\
$\Gamma_2$ & 1 & 1 &  1 & -1 & -1 &  1 &  1 &  1 & -1 & -1 \\
$\Gamma_3$ & 1 & 1 & -1 & -1 &  1 &  1 &  1 & -1 & -1 &  1 \\
$\Gamma_4$ & 1 & 1 &  1 &  1 &  1 & -1 & -1 & -1 & -1 & -1 \\
$\Gamma_5$ & 1 & 1 & -1 &  1 & -1 & -1 & -1 &  1 & -1 &  1 \\
$\Gamma_6$ & 1 & 1 &  1 & -1 & -1 & -1 & -1 & -1 &  1 &  1 \\
$\Gamma_7$ & 1 & 1 & -1 & -1 &  1 & -1 & -1 &  1 &  1 & -1 \\
\hline
$\Gamma_8$ & 2 & -2 &  0 & 0 &  0 &  2 & -2 &  0 &  0 &  0 \\
$\Gamma_9$ & 2 & -2 &  0 & 0 &  0 & -2 &  2 &  0 &  0 &  0 \\
\hline\hline
 \end{tabular}
\end{center}
\end{table}

\begin{table}
 \caption{Character Table for the $C_{2v}$ double group. The first line labels the symmetry operations in short notation. For ribbon C we have 
 $C_2 = C_2(y)$, $M = M(xy)$, and $M'=M(yz)$. Barred and unbarred operations refer to $4\pi$ and $2\pi$ rotations on spin space. The first set, from $\Gamma'_0$ to $\Gamma'_3$, are single group irreducible representations, and $\Gamma'_4$ is the double group irreducible representation.}
 \label{tab:C2v}
\begin{center}
\begin{tabular}{l | c c c c c}
\hline\hline
 & $E$ & $\bar{E}$ & $\begin{array}{c}C_2 \\ \bar{C}_2\end{array}$ & $\begin{array}{c}M \\ \bar{M}\end{array}$ & $\begin{array}{c}M' \\ \bar{M}'\end{array}$\\
\hline
$\Gamma'_0$ & 1 & 1 &  1 &  1 &  1 \\
$\Gamma'_1$ & 1 & 1 & -1 &  1 & -1 \\
$\Gamma'_2$ & 1 & 1 &  1 & -1 & -1 \\
$\Gamma'_3$ & 1 & 1 & -1 & -1 &  1 \\
\hline
$\Gamma'_4$ & 2 & -2 &  0 & 0 &  0 \\
\hline\hline
 \end{tabular}
\end{center}
\end{table}

\begin{table*}
 \caption{Character Table for the $D_{2h}^{NS}$ nonsymmorphic double group. The first line labels the symmetry operations in short notation, where $\mathcal{O}_0 = \{\mathcal{O},0\}$ and $\mathcal{O}_\chi = \{\mathcal{O},\chi\}$ in the nonsymmorphic element representation, $\chi$ is the translation by half a unit cell, and the point group operations follow the ones of Ribbon D in Table \ref{tab:D2h}. The first set, from $\Gamma''_0$ to $\Gamma''_9$, are single group irreducible representations (IRREPs), and from $\Gamma''_{10}$ to $\Gamma''_{12}$ are the double group IRREPs. The highlighted IRREPs are the ones allowed for the $\bm{k} = X$ TRIM due to Bloch periodicity, i.e. $E_\chi = -E_0$.}
 \label{tab:D2hNS}
\begin{center}
\begin{tabular}{l !{\vrule width -4pt} | !{\vrule width -4pt} c !{\vrule width -4pt} c !{\vrule width -4pt} c!{\vrule width -4pt} c!{\vrule width -4pt} c!{\vrule width -4pt} c !{\vrule width -4pt} c !{\vrule width -4pt} c !{\vrule width -4pt} c !{\vrule width -4pt} c !{\vrule width -4pt} c !{\vrule width -4pt} c !{\vrule width -4pt} c !{\vrule width -4pt} c}
\hline\hline
 & $E_0$ & $\bar{E}_0$ & $\begin{array}{c}M''_\chi\\\bar{M}''_\chi\\M''_{3\chi}\\\bar{M}''_{3\chi}\end{array}$ & $E_{2\chi}$ & $\bar{E}_{2\chi}$ & $\begin{array}{c}M'\chi\\\bar{M}'_\chi\\M'_{3\chi}\\\bar{M}'_{3\chi}\end{array}$ & $\begin{array}{c}C_{2,0}\\C_{2,2\chi}\end{array}$ & $\begin{array}{c}\bar{C}_{2,0}\\\bar{C}_{2,2\chi}\end{array}$ & $\begin{array}{c}M_0\\\bar{M}_0\end{array}$ & $\begin{array}{c}C''_\chi\\\bar{C}''_\chi\\C''_{3\chi}\\\bar{C}''_{3\chi}\end{array}$ & $\begin{array}{c}M_{2\chi}\\\bar{M}_{2\chi}\end{array}$ & $\begin{array}{c}C'_\chi\\\bar{C}'_\chi\\C'_{3\chi}\\\bar{C}'_{3\chi}\end{array}$ & $\begin{array}{c}I_0\\I_{2\chi}\end{array}$ & $\begin{array}{c}\bar{I}_0\\\bar{I}_{2\chi}\end{array}$\\
\hline
$\Gamma''_0$ & 1 & 1 &  1 &  1 &  1 &  1 &  1 &  1 &  1 &  1 & 1 &  1 &  1 &  1\\
$\Gamma''_1$ & 1 & 1 &  1 &  1 &  1 & -1 & -1 & -1 &  1 &  1 & 1 & -1 & -1 & -1\\
$\Gamma''_2$ & 1 & 1 & -1 &  1 &  1 &  1 & -1 & -1 &  1 & -1 & 1 &  1 & -1 & -1\\
$\Gamma''_3$ & 1 & 1 & -1 &  1 &  1 & -1 &  1 &  1 &  1 & -1 & 1 & -1 &  1 &  1\\
\rowcolor{gray!25}
$\Gamma''_4$ & 2 & 2 & 0 & -2 & -2 & 0 & 0 & 0 & 2 & 0 & -2 & 0 & 0 & 0\\
$\Gamma''_5$ & 1 & 1 & 1 & 1 & 1 & 1 & 1 & 1 & -1 & -1 & -1 & -1 & -1 & -1\\
$\Gamma''_6$ & 1 & 1 & 1 & 1 & 1 & -1 & -1 & -1 & -1 & -1 & -1 & 1 & 1 & 1\\
$\Gamma''_7$ & 1 & 1 & -1 & 1 & 1 & 1 & -1 & -1 & -1 & 1 & -1 & -1 & 1 & 1\\
$\Gamma''_8$ & 1 & 1 & -1 & 1 & 1 & -1 & 1 & 1 & -1 & 1 & -1 & 1 & -1 & -1\\
\rowcolor{gray!25}
$\Gamma''_9$ & 2 & 2 & 0 & -2 & -2 & 0 & 0 & 0 & -2 & 0 & 2 & 0 & 0 & 0\\
\hline\hline
$\Gamma''_{10}$ & 2 & -2 & 0 & 2 & -2 & 0 & 0 & 0 & 0 & 0 & 0 & 0 & 2 & -2\\
$\Gamma''_{11}$ & 2 & -2 & 0 & 2 & -2 & 0 & 0 & 0 & 0 & 0 & 0 & 0 & -2 & 2\\
\rowcolor{gray!25}
              & 2 & -2 & 0 & -2 & 2 & 0 & -2 & 2 & 0 & 0 & 0 & 0 & 0 & 0\\
\rowcolor{gray!25}
\multirow{-2}{*}{$\Gamma''_{12}\Big\{$}
              & 2 & -2 & 0 & -2 & 2 & 0 & 2 & -2 & 0 & 0 & 0 & 0 & 0 & 0\\
\hline\hline
\end{tabular}
\end{center}
\end{table*}

Since we are dealing with a spinfull system, the eigenstates must belong to the double group irreducible representations (IRREPs). For the monolayer sheet, and ribbons A and B the relevant IRREPs are $\Gamma_8$ + $\Gamma_9$ in Table \ref{tab:D2h}, both two-dimensional. Consequently, the band structure of ribbons A and B can only have twofold degeneracies at $\bm{k}=X$. This is clearly seen in Fig.~\ref{fig:BSABCDE}(A-B) on the minigap between edge states. Each line is doubly degenerate, and the minigap can only close asymptotically for wide ribbons to form fourfold degeneracies. In other words, the gapless crossing point of the edge states is not protected for narrow ribbons.

For ribbon C, all spinfull states belong to the $\Gamma'_4$ IRREP of Table \ref{tab:C2v}, which is again two-dimensional, and ribbon C can only have doubly degenerate states at $\bm{k}=X$. However, here the edge states from opposite sites see different atomic terminations, and their linear dispersions occur at different energies, see Fig.~\ref{fig:BSABCDE}(C). Consequently, the states from opposite edges that would hybridize for narrow ribbons are already energy splitted due to the boundary conditions. This implies that the gapless edge state dispersions seen in Fig.~\ref{fig:BSABCDE}(C) are robust even for narrow ribbons, as long as they remain energy splitted.

Ribbon D also has two-dimensional IRREPs only, $\Gamma_8$ and $\Gamma_9$ of Table \ref{tab:D2h}, and once again, it can only have twofold degeneracies at $\bm{k}=X$. In Fig.~\ref{fig:BSABCDE}(D) each edge state is doubly degenerate, therefore, at the crossing at $\bm{k}=X$ a hybridization gap opens to maintain a two-fold degeneracy. This gap closes asymptotically with the ribbon width to form four-fold degeneracies.

\subsection{Nonsymmorphic Ribbon E}

The nonsymmorphic $D_{2h}^{NS}$ group of ribbon E is shown in Table \ref{tab:D2hNS}. To obtain this character table we have to deal with the nonsymmorphic multiplications defined as \cite{car1976optimal,koster2012space,dresselhaus2007group}

\begin{align}
 \{\alpha | \bm{\chi}_1 \}\{\beta | \bm{\chi}_2 \} =& \{\alpha\beta | \alpha\bm{\chi}_2+\bm{\chi}_1 \},\\
 \{\alpha | \bm{\chi}_1 \}^{-1} =& \{\alpha^{-1} | - \alpha^{-1}\bm{\chi}_1 \},
\end{align}
where $\alpha$ and $\beta$ are regular point group symmetry operations (e.g., inversion, mirrors, rotations), and $\bm{\chi}$ are lattice translations. A crystal is said nonsymmorphic if it is not possible to find an origin for the coordinates such that all symmetry operations can be written using only the regular point group operations and primitive translations. Instead, nonsymmorphic crystals require symmetry operations that contain translations $\bm{\chi}$ by less than a unit cell. This is the case of ribbon E, see Eqs.~\eqref{eq:NSelements}.

Reference \cite{car1976optimal,dresselhaus2007group} shows that for a nonsymmorphic crystal, one typically needs a factor group that is usually twice the size of the wave-vector group in order to close the multiplication table (not shown here). Indeed in our case we find that one needs the symmetry elements of Eqs.~\eqref{eq:NSelements}, plus equivalent ones with an extra translation by $\{0, 2\chi\}$, where $2\chi$ is one unit cell along $\hat{x}$. With the multiplication table at hand, we identify the symmetry classes and construct the Character Table \ref{tab:D2hNS} following the usual procedure\cite{voon2009kp, dresselhaus2007group, car1976optimal, koster2012space, FazzioTeoGrupos}. However, not all IRREPs are compatible with the eigenstates of a given point $\bm{k}$. The matrix representation of a nonsymmorphic symmetry element $\{\alpha | \bm{\chi} \}$ is $D(\{\alpha | \bm{\chi} \}) = e^{i\bm{k}\cdot\bm{\chi}}D(\alpha)$, where $D(\alpha)$ is the matrix representation of the point group element $\alpha$. Take now the elements $\{E | 0\}$ and $\{E | 2\chi\}$,

\begin{align}
 D(\{E|0\}) &= D(E),\\
 D(\{E|2\chi\}) &= e^{i k_x 2\chi} D(E) = -D(E),
\end{align}
where $e^{i k_x 2\chi} = -1$ for $\bm{k} = X$ refers to the Bloch periodicity at the X TRIM. Consequently, only the IRREPs that satisfy this condition are compatible with the eigenstates at X. These IRREPs are highlighted in grey in Table \ref{tab:D2hNS}.

For the spinful system, only the $\Gamma''_{12}$ double group IRREPs are compatible with the Bloch periodicity. From the Frobenius-Schur test, the Herring rules \cite{dresselhaus2007group} state that this pair of IRREPs stick together to form a four-dimensional IRREP $\Gamma''_{12}$ (given by the sum of both lines) in order to satisfy time-reversal symmetry. Therefore the states of ribbon E are expected to show four-fold degeneracy at $\bm{k} = X$, as seen in Fig.~\ref{fig:BSABCDE}(E). This degeneracy requirement determinies that the Dirac cone in ribbon E must remain gapless even for narrow ribbons.

\bibliography{pbse}

\bibliography{pbse}

\end{document}